\documentstyle[twoside,fleqn,espcrc2,epsf]{article}

\newcommand{\non}{\nonumber}

\title{Approximate calculation of corrections at NLO and NNLO.
\thanks{
Also supported by
the Secretariat for Research and Technology of Greece and by the Natural Sciences
and Engineering Research Council of Canada.}}

\author{
A. P. Contogouris \address{Nuclear and Particle Physics,
University of Athens, Athens 15771, Greece}$^{,}$
\address{Department of Physics, McGill University, Montreal
H3A2T8, Canada} and
G. Grispos $^{a}$}
      
\begin{document}
\pagestyle{empty}

\begin{abstract}
For processes involving structure functions and/or fragmentation functions, arguments
that there is a part that dominates the NLO corrections are briefly reviewed. The arguments
are tested against more recent NLO and in particular NNLO calculations. 
\end{abstract}

\maketitle

\section{THE DOMINANT PART AND ITS IMPLICATIONS}

For many unpolarized and polarized inclusive reactions calculations have now been carried in
next-to-leading order (NLO) in the running coupling $\alpha_s (Q)$ and in a few of them in
next-to-next-to-leading order (NNLO). All analytic NNLO results are very complicated and the same holds
for the NLO ones when the leading order (LO, Born) subprocess corresponds to a 4-point function.

For processes involving structure functions and/or fragmentation functions, in \cite{1} it was argued that
there is a part that dominates the NLO; and this was used to explain the fact that, in a number of the
then existing NLO calculations, plotted against the proper kinematic variable, in a wide range, the
cross section was almost a constant multiple of the Born.

Here we extend these considerations to a number of NLO calculations carried in the meantime,
in particular on polarized reactions, as well as to the existing NNLO ones.

To briefly review the essential ideas of \cite{1}, consider the NLO contribution of the subprosses
$a(p_1)+b(p_2) \rightarrow \gamma (q)+d$ to the large-$p_T$ process $A+B \rightarrow \gamma +X$:
\begin{eqnarray}
E \frac{d \sigma}{d^3 p}= \frac{\alpha_s (\mu )}{\pi } \int dx_a dx_b
F_{a/A} \left( x_a, M\right) \non \\
F_{b/B} \left( x_b, M\right) \bigg[ \hat{\sigma} _B \delta \left(1+\frac{t+u}{s}\right)+ \non \\
\frac{\alpha_s (\mu )}{\pi } f \theta \left(1+\frac{t+u}{s}\right) +cross. \,term \bigg]
\end{eqnarray}
where $F_{a/A}$, $F_{b/B} $ are parton momentum distributions, $\mu$ and $M$ are of $O(p_T)$,
\begin{eqnarray}
s=\left( p_1 +p_2 \right) ^2 ,\quad t=\left( q -p_1 \right) ^2 ,\quad u=\left( q -p_2 \right) ^2 \non
\end{eqnarray}
and $\sigma_B$ and $f$ are functions of $s$, $t$, $u$ corresponding to the Born and the higher
order correction (HOC). Introducing the dimensionless variables
\begin{eqnarray}
v=1 +t/s ,\quad w=-u/ \left( s+t \right)
\end{eqnarray}
($s+t+u=sv(1-w)$), the HOC have the following overall structure:
\begin{eqnarray}
f \left( v,w\right) =f_s \left( v,w\right) +f_h \left( v,w\right) \non
\end{eqnarray}
where
\begin{eqnarray}
f_s \left( v,w \right) = a_1 (v) \delta (1-w) +b_1 (v) \frac{1}{(1-w)_+} + \non\\
c(v) \left( \frac{ln (1-w) }{1-w} \right) _+ + \bigg( a_2 (v) \delta (1-w) + \non\\
b_2 (v) \frac{1}{(1-w)_+} \bigg) ln \frac{s}{M^2}
\end{eqnarray}
$f_h \left( v,w\right) $ contains no distributions and, in general, is very complicated.

Now denote by $\sigma_s$ and $\sigma_h$ the contributions of $f_s$ and $f_h$ to $E d \sigma / d^3 p$
and consider the ratio
\begin{eqnarray}
L=\sigma_h / (\sigma_s +\sigma_h) ;
\end{eqnarray}
then, for fixed total c.m. energy $\sqrt{S}$, as $p_T$ (or $x_T \equiv 2p_T/ \sqrt{S}$) increases, $L$ decreases.

To see the reason, consider a plot of $x_b$ vs $x_a$ (Fig. 1 of [1a]). The integration region in (1) is bounded
by $w=1$, $x_a=1$ and $x_b=1$. Now, for $x$ not too small, $F_{a/A} (x, M)$ behaves like $(1-x)^n$; with
$A=$proton, $n$ is fairly large ($\geq 3$); also due to scale violations, $n$ increases as $p_T$ increases.
Then contributions arising from the region away from $w=1$ are supressed by powers of $1-x_a$ and/or $1-x_b$.
Now, in $f_s$, the terms $\sim \delta (1-w)$ contribute at $w=1$ (and so does $\hat{\sigma}_B$) whereas the
rest give a contribution increasing as $w\rightarrow 1$. On the other hand, the multitude of terms of $f_h$
contribute more or less uniformly in the integration region $\theta(1-w)$ and their contribution $\sigma _h$
is suppressed. As $x_T$ increases at fixed $S$, the integration region shrinks towards $x_a =x_b =1$
(Fig. 1 of [1a]) and the suppression of $\sigma _h$ increases.
   
The mechanism is tested by writing the distributions in the form [1a]:
\begin{eqnarray}
F_{a/A} (x, M)=F_{b/B} (x, M)=(1-x)^N
\end{eqnarray}
and choosing a fictitious $N>>n$ or $0<N<<n$. Then the ratio $L$ in the first case decreases  faster, in the
second slower.

Neglecting $f_h (v,w)$ and with the rough approximations $1/(1-w)_+ \sim \delta (1-w)$,
$(ln(1-w)/(1-w))_+ \sim \delta (1-w)$ we obtain in (1): $f \approx  \delta (1-w)$ resulting in $E d \sigma / d^3 p$
of roughly the same shape as $E d \sigma_{Born} / d^3 p$.

At NLO the Bremsstrahlung (Brems) contributions to $f_s$ are determined via simple formulae \cite{1}:
E.g. for $gq\rightarrow \gamma q$ the Brems contributions arise from products of two graphs
$gq\rightarrow \gamma qg$. If in both graphs the emitted $g$ arises from initial partons (g or q),
the contribution in $n=4-2\varepsilon $ dimensions is
\begin{eqnarray}
\frac{d\sigma _{init}}{dvdw} \sim T_0 ^{(gq)} (v,\varepsilon ) N_c \left( -\frac{2}{\varepsilon }\right)
\left( \frac{v}{1-v} \right) ^{-\varepsilon } \non\\
(1-w)^{-1-2\varepsilon } \left( 1+\varepsilon ^2 \frac{\pi ^2}{6} \right)
\end{eqnarray}
where $T_0 ^{(gq)} (v,\varepsilon ) $ is essentially the Born cross section in $n$ dimensions. If in
at least on of the graphs the emitted $g$ arises from the final parton ($q$), then
\begin{eqnarray}
\frac{d\sigma _{fin}}{dvdw} \sim T_0 ^{(gq)} (v,\varepsilon ) C_F v^{-\varepsilon }
(1-w)^{-1-\varepsilon } \tilde P _{qq} (\varepsilon )
\end{eqnarray}
where
\begin{eqnarray}
\tilde P _{qq} (\varepsilon )=\frac{\Gamma(1-2\varepsilon )}{\Gamma ^2 (1- \varepsilon )}
\int _0 ^1 y^{-\varepsilon } (1-y) ^{-\varepsilon } P_{qq} (y,\varepsilon ) \non
\end{eqnarray}
and $P_{qq} (y,\varepsilon ) =2/(1-y)-1-y-\varepsilon (1-y)$, the split function in $n$ dimensions
($y<1$). Expanding
\begin{eqnarray}
(1-w)^{-1-\varepsilon }=-\frac{1}{\varepsilon } \delta (1-w) +\frac{1}{(1-w)_+} \non\\
-\varepsilon \left( \frac{ln(1-w)}{1-w} \right)_+ +O(\varepsilon ^2) \non
\end{eqnarray}
as well as $(v/(1-v))^{-\varepsilon }$ and  $v^{-\varepsilon }$ in powers of $\varepsilon $ we determine the
contributions. The singular terms $\sim 1/\varepsilon ^2$ and $1/\varepsilon $ cancel by adding the loop
contributions and proper counterterms.

\section{FURTHER NLO CALCULATIONS}

In addition to the examples presented in Refs \cite{1}, the following are some NLO studies supporting the
ideas of Sect. 1:
\begin{itemize}
\item[(a)]
Heavy quark $Q$ production in $p \bar p$ collisions \cite{2}. The cross sections $d\sigma /dydp_T ^2$
versus $p_T$ of $Q$ for several rapidities $y$ and for $m_Q =5$, 40 and 80 $GeV$ at $\sqrt{S}=0.63$
and 1.8 $TeV$ are a constant multiple of the LO one (Figs 7-12). See also \cite{3} Fig. 10.15. In all the
cases the verification is striking.
\item[(b)]
Large $p_T$ $W$ and $Z$ production in $p \bar p$ collisions \cite{4}. At $\sqrt{S}=0.63$ and 1.8 $TeV$,
for $p_T \geq 80$ $GeV$ the cross sections $d\sigma /dp_T ^2$ are also almost a constant multiple of
the LO (Figs 7 and 8).
\end{itemize}

Regarding NLO results for polarized reactions we mention the following:  
\begin{itemize}
\item[(a)]
Polarized deep inelastic Compton scattering \cite{5}, in particular the contribution of the subprocess
$\vec \gamma \vec p \rightarrow \gamma q$ to large $p_T$ $\vec \gamma \vec q \rightarrow \gamma +X$.
At $\sqrt{S}=27$ and 170 $GeV$, for $x_T \geq 0.15$, it is $L<0.28$ and for sufficiently large $x_T$, $L$
decreases as $x_T\rightarrow 1$ (Ref. 5, Fig. 4). Also, denoting by $\sigma ^{(k)}$ the $O(\alpha _s ^k )$, $k=0,1$,
contributions of $\vec \gamma \vec q \rightarrow \gamma q$ to $E d \sigma / d^3 p$, for $0.2\leq x_T \leq 0.8$
the factor $K_{\gamma q}=(\sigma ^{(0)} +\sigma ^{(1)})/\sigma ^{(0)}$ is found to differ little from a constant.
\item[(b)]
Large $p_T$ direct $\gamma $ production in longitudinally polarized hadron collisions [6, 7]. Here of interest
are the $O(\alpha _s ^k )$, $k=1,2$, of the subprocess $\vec g \vec q \rightarrow \gamma q$. As $x_T$
increases, the ratio $-\sigma _h /\sigma _s $ steadily decreases (Ref. 6, Fig. 10). The factor
$K_{gq}=(\sigma ^{(1)} +\sigma ^{(2)})/\sigma ^{(1)}$ is not constant, but increases moderately (Fig. 2).
\item[(c)]
Lepton pair production by transversely polarized hadrons [8, 9]. At fixed $S$, with increasing
$\sqrt{\tau }= M_{l^- l^+}/\sqrt{S}$, the ratio $\sigma _h /\sigma _s $ is found again to decrease (Ref. 8, Fig. 3).
Again, the $K$-factor is not constant, but increases moderately (Ref. 8, Fig. 1). 
\end{itemize}

The considerations of Sect. 1 explain also the following fact: Taking as example large $p_T$
$\vec p \vec p \rightarrow \gamma +X$, at NLO, apart from the HOC of the dominant subprocess
$\vec g \vec q \rightarrow \gamma q$, there are contributions from the extra subprocesses
$\vec g \vec g \rightarrow q \bar q \gamma $ and $\vec q \vec q \rightarrow qq \gamma $. In general, these are
found to be small (Ref. 6, Figs 3, 4 and 5). The reason is that the extra subprocesses possess no terms involving
distributions (no loops and vanishing contributions of the type (6) and (7)).

\section{NNLO CALCULATIONS}

NNLO calculations have been carried for Drell-Yan (DY) production of lepton pairs, $W^{\pm }$ and $Z$, and for
the deep inelastic (DIS) structure functions $F_j (x, Q^2 )$, $j=1,2$ and $L$. Now the parts involving distributions
contain also terms of the type $(ln^i (1-w)/(1-w))_+ $, with $i=2$ and 3 and $w$ a proper dimensionless variable.
Calculations are carried using the set $S- \overline{MS}$ of \cite{10}.

Beginning with DY, we are interested in the process $h_1 h_2 \rightarrow \gamma ^* +X \rightarrow l^+ l^- +X$ and
to the cross section
\begin{eqnarray}
d\sigma (\tau , S) /dQ^2 \equiv \sigma (\tau , S) 
\end{eqnarray}
where $\tau =Q^2 /S$ with $\sqrt{S}$ the total c.m. energy of the initial hadrons $h_1$, $h_2$ and $\sqrt{Q^2 }$
the $\gamma ^* $ mass [11, 12]. Here we deal with the subprocess $q +\bar q\rightarrow \gamma ^*$ and its NLO and
NNLO corrections \cite{11}. For DY, $w\sim \tau $.

Denote by $\sigma ^{(k)} (\tau , S)$, $k=0,1,2$, the $O(\alpha _s ^k )$ part of $\sigma (\tau , S)$, by $\sigma _s ^{(k)}$
the part of $\sigma ^{(k)}$ arising from distributions and by $\sigma _h ^{(k)}$ the rest. Defining
\begin{eqnarray}
L^{(k)} (\tau , S) = \sigma _h ^{(k)} (\tau , S) /  \sigma ^{(k)} (\tau , S) 
\end{eqnarray}
Fig. 1 shows $L^{(k)}$, $k=1,2$, as functions of $\tau $ for $\sqrt{S}=20$ $GeV$. Clearly, for $\tau >0.3$: $L^{(1)} \leq 0.16$
and $L^{(2)} \leq 0.4$.

It is of interest also to see the percentage of $\sigma _h ^{(k)}$ of the total cross section determined up to $O(\alpha _s ^k )$.
Fig. 1 also shows the ratios $\sigma _h ^{(1)}/(\sigma ^{(0)} +\sigma ^{(1)} )$ and
$\sigma _h ^{(2)}/(\sigma ^{(0)} +\sigma ^{(1)} +\sigma ^{(2)} )$ for the same $\sqrt{S}$; clearly, for $\tau \geq 0.2$
both ratios are less than 0.1.  

Now we turn to DIS [13, 14] and present results for the contribution to the structure function $F_2 (x,Q^2)$ of the d-valence
quark distribution. We will deal with the subprocess $q+\gamma ^* \rightarrow q$ and its NLO and NNLO corrections \cite{14}.
For DIS, $w\sim x$.

Denote by $F^{(k)} (x,Q^2 )$, $k=0,1,2$, the $O(\alpha _s ^k )$ contribution, by $F_s ^{(k)}$
the part of $F^{(k)}$ arising from distributions and by $F_h ^{(k)}$ the rest. Defining
\begin{eqnarray}
L^{(k)} (x,Q^2 ) = F_h ^{(k)} (x,Q^2 ) /  F^{(k)} (x,Q^2 )
\end{eqnarray}
Fig. 2 presents $L^{(k)}(x,Q^2 )$, $k=1,2$, as functions of $x$ for $\sqrt{Q^2}=5$ $GeV$. Now, for $x\leq 0.5$ $L^{(1)}$ is not
small, but this is due to the fact that $F_s ^{(1)}$ changes sign and $F_h ^{(1)}$ stays $>0$, so at $x\approx 0.3$ $F^{(1)}$
vanishes. On the other hand, at $x\geq 0.3$, $L ^{(2)}$ is less than 0.2.

Fig. 2 also shows the ratios $F_h ^{(1)}/(F ^{(0)} +F ^{(1)} )$ and
$F_h ^{(2)}/(F^{(0)} +F^{(1)} +F ^{(2)} )$ for the same $Q^2$; for $x \geq 0.3$ both ratios are less than 0.08.

The effect of neglecting $\sigma _h ^{(k)}$ in DY or $F_h ^{(k)}$ in DIS is shown in Fig.  3. In DY, denoting  
\begin{eqnarray}
K_s=(\sigma ^{(0)}+\sigma _s ^{(1)} +\sigma _s ^{(2)} )/\sigma ^{(0)} \non\\
K=(\sigma^{(0)}+\sigma ^{(1)} +\sigma ^{(2)})/\sigma ^{(0)}  
\end{eqnarray}
we show $K_s (K)$ by solid (dashed) line at $\sqrt{S}=20$ $GeV$ (upper part). Clearly, as $\tau \rightarrow 1$, 
$K_s \rightarrow K$, and for $\tau > 0.3$ the error is less than $12\%$. 
In DIS, denoting by $K_s$ and $K$ the
$K$-factors of (11) with $\sigma ^{(k)}$ replaced by $F ^{(k)}$, we show $K_s$ and $K$ at $\sqrt{Q^2}=5$ $GeV$
(lower part). Again, as $x \rightarrow 1$, $K_s \rightarrow K$. Now, in spite of the fact that $L^{(k)}$ is, in general,
not small, $K_s$ differs from $K$ even less. The reason is that the NLO and NNLO corrections are smaller than
in DY, and so are $F_s ^{(k)} /F^{(0)}$.

\section{CONCLUSIONS}

The above discussion and examples show that for processes involving structure and/or fragmentation functions,
for not too small values of a proper kinematic variable ($x_T$ for large-$p_T$ reactions, $\tau $ for DY, $x$ for DIS),
with reasonable accuracy one can retain only the part of the differential cross section arising from distributions
(dominant part).

\begin{figure}
\centering
\mbox{\epsfxsize=60mm\epsfysize=50mm\epsffile{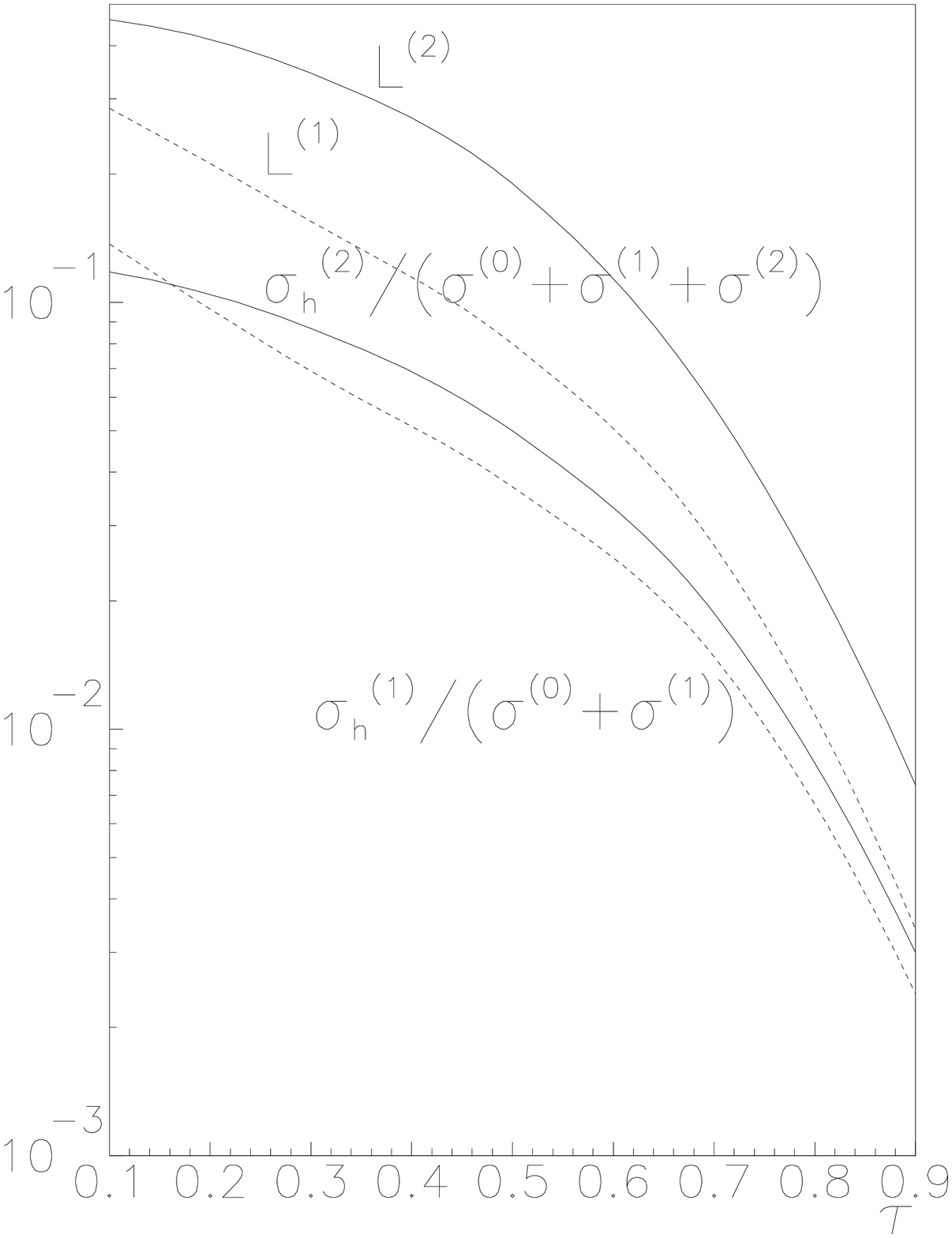}}
\vskip -.35in
\begin{center}
Fig. 1
\end{center}
\end{figure}
\begin{figure}
\centering
\mbox{\epsfxsize=60mm\epsfysize=50mm\epsffile{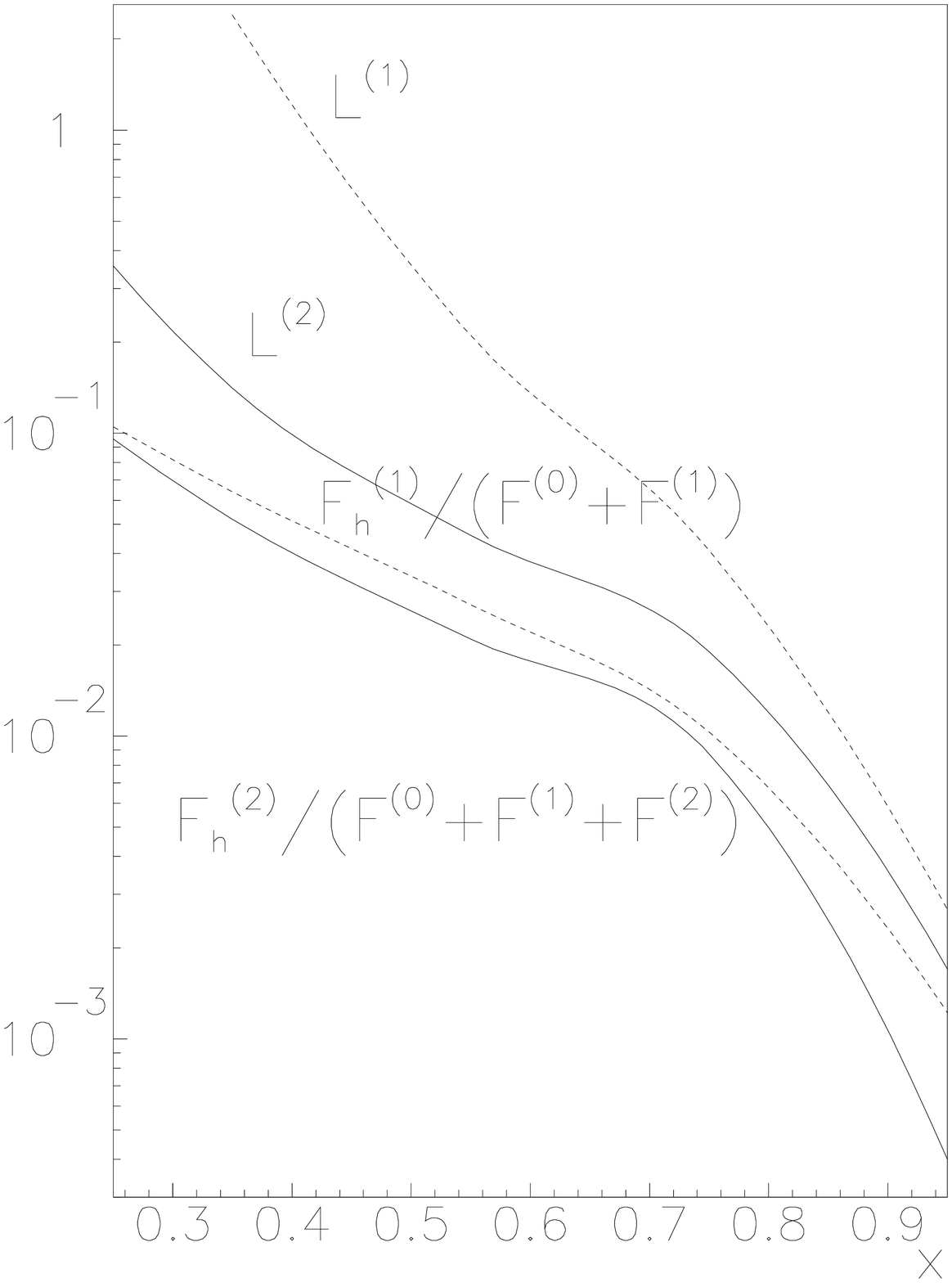}}
\vskip -.35in
\begin{center}
Fig. 2
\end{center}
\end{figure}  
\begin{figure}
\centering
\mbox{\epsfxsize=60mm\epsfysize=50mm\epsffile{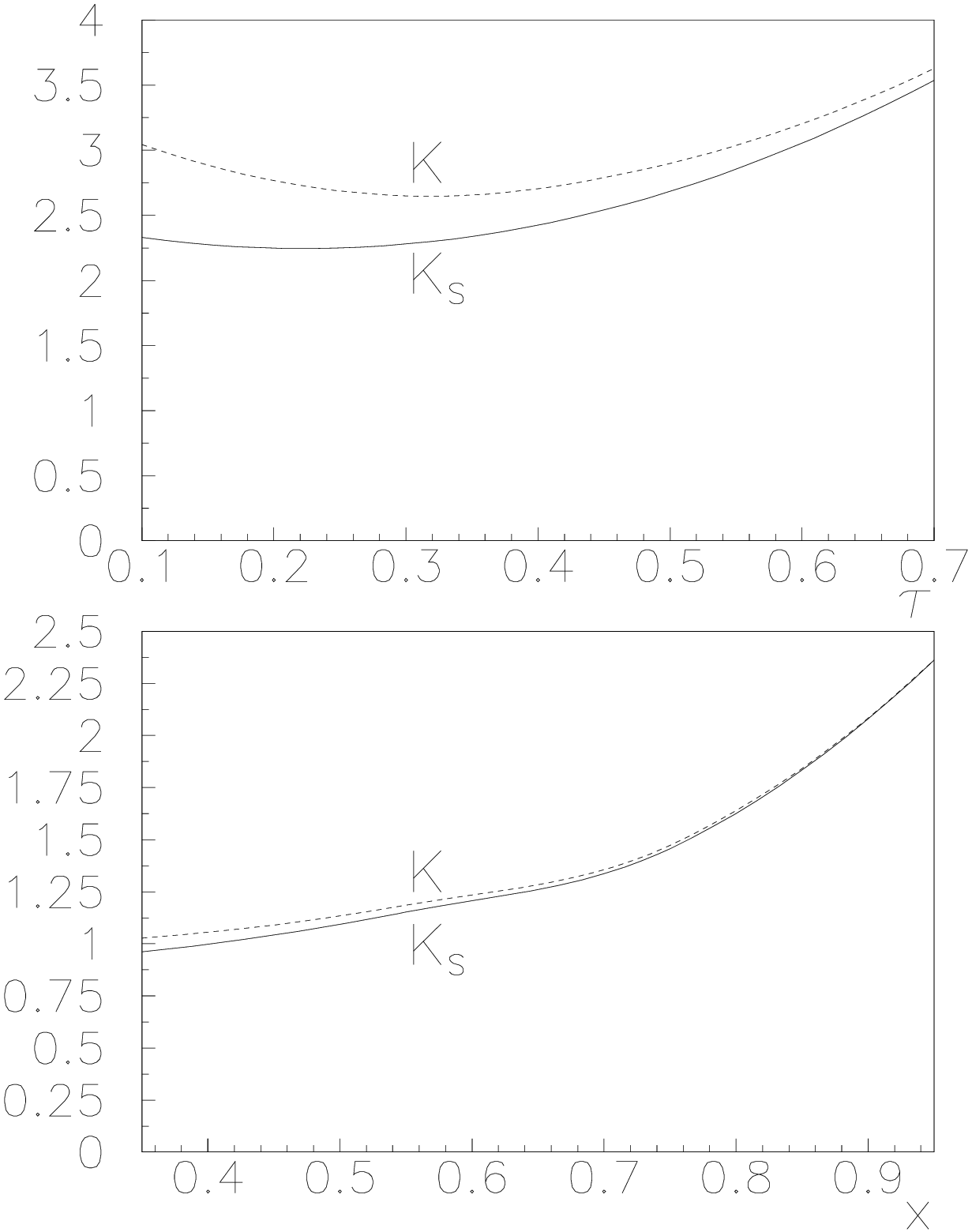}}
\vskip -.35in
\begin{center}
Fig. 3
\end{center}
\end{figure}  


\begin{thebibliography}{9}
\bibitem{1}
(a) A. P. Contogouris, N. Mebarki, S. Papadopoulos,
Intern. J. Mod. Phys. A5 (1990) 1951.
(b) A. P. Contogouris, S. Papadopoulos,
Mod. Phys. Lett. A5 (1990) 901.
\bibitem{2}
P. Nason, S. Dawson, R. Ellis,
Nucl. Phys. B327 (1989) 49.
\bibitem{3}
R. Ellis, W. Stirling, B. Weber,
"QCD and Collider Physics" (Cambridge Univ. Press, 1996).
\bibitem{4}
R. Gonsalves, J. Pawlowski, C. F. Wai,
Phys. Rev. D40 (1989) 2245.
\bibitem{5}
A. P. Contogouris, S. Papadopoulos, F. V. Tkachov,
ibid D46 (1992) 2846.
\bibitem{6}
A. P. Contogouris, B. Kamal, Z. Merebashvili, F. V. Tkachov,
ibid D48 (1993) 4092; D54 (1996) 701 (E).
\bibitem{7}
L. Gordon, W. Vogelsang,
ibid D49 (1994) 70.
\bibitem{8}
A. P. Contogouris, B. Kamal, Z. Merebashvili,
Phys. Lett. B337 (1994) 169.
\bibitem{9}
W. Vogelsang, A. Weber,
Phys. Rev. D48 (1993) 2073.
\bibitem{10}
J. Morfin, W. K. Tung,
Z. Phys. C52 (1991) 13.
\bibitem{11}
T. Matsuura, S. van der Marck, W. van Neerven,
Phys. Lett. B211 (1988) 171; Nucl. Phys. B319 (1989) 570.
\bibitem{12}
P. Rijken, W. van Neerven,
Phys. Rev. D51 (1995) 44.
\bibitem{13}
E. Zijlstra, W. van Neerven,
Phys. Lett. B273 (1991) 476; Nucl. Phys. B383 (1992) 525.
\bibitem{14}
W. van Neerven, E. Zijlstra, 
Phys. Lett. B272 (1991) 127.
\end{thebibliography}
\end{document}